\documentclass[eqsecnum,aps,epsfig,nofootinbib]{revtex4}
\begin{document}

\def\beq{\begin{equation}}
\def\endeq{\end{equation}}
\def\bea{\begin{eqnarray}}
\def\endea{\end{eqnarray}}

\title{Perfect mirrors and the self-accelerating box paradox}

\author{Donald Marolf}

\affiliation{
Physics Department, Syracuse University, Syracuse, NY 13244.}

\author{Rafael D. Sorkin}
\affiliation{
Physics Department, Syracuse University, Syracuse, NY 13244.}

\date{\today}

\begin{abstract}
We consider the question raised by Unruh and Wald of whether
mirrored boxes can self-accelerate in flat spacetime (the
``self-accelerating box paradox'').
{}From the point of view of the box, which perceives the acceleration as
an impressed gravitational field, this is equivalent to asking
whether the box can be supported by the buoyant force arising from its
immersion in a perceived bath of thermal (Unruh) radiation.
The perfect mirrors we study are of the type that rely on light internal
degrees of freedom which adjust to and reflect impinging radiation.
We suggest that a minimum of one internal mirror degree of freedom is
required for each bulk field degree of freedom reflected.  
A short calculation then shows that such mirrors necessarily absorb
enough heat from the thermal bath that their increased mass
prevents them from floating on the thermal radiation.  
For this type of mirror the paradox is therefore resolved.
We also observe that this failure of boxes to ``float'' invalidates one
of the assumptions going into the Unruh-Wald analysis of entropy
balances involving boxes lowered adiabatically toward black holes.  
Nevertheless, their broad argument can be maintained until the box
reaches a new regime in which box-antibox pairs dominate
over massless fields as contributions to thermal radiation.  
\end{abstract}
\maketitle


\section{Introduction}

\label{intro}
Black hole entropy, Hawking radiation, and the generalized second law of
thermodynamics are central foci of modern research in quantum gravity.
Indeed, although the details may differ, the existence of these
phenomena is one of the few points on which the various approaches to
quantum gravity all seem to agree, including
the stringy approach \cite{SV,Arev},
the causal set approach \cite{Dou,Dou-rds}, and
the loop approach \cite{ABCK,ABK}.  
Many researchers believe that these phenomena are truly fundamental and
that their investigation will reveal further deep laws of nature.
Certainly, black hole thermodynamics has provided a fascinating and
varied arena for nearly thirty years of research.

One of the most intriguing effects to emerge from this discussion is the
experience of thermal radiation by an accelerating object in flat and
empty spacetime \cite{U}.  This Unruh radiation has a temperature $T$
proportional to the acceleration $a$, and it affects accelerating
objects as one would expect, setting off particle detectors and exerting
pressure.  In a discussion \cite{UW} of a potential violation (suggested
by Bekenstein \cite{Bek73}) of the generalized second law of
thermodynamics, Unruh and Wald pointed out that the pressure on an
accelerating box will vary with position, and that the pressure
difference across the box must grow quickly with the acceleration.  For
a perfectly reflecting box of unchanging mass there would therefore be a
(box-dependent) critical acceleration $a_c$ at which the pressure
difference would be big enough to maintain the acceleration without the
need for any additional force.  
{}From the point of view of the box,
it experiences a gravitational field of strength $a_c$, but the
radiation fluid is sufficiently dense that the mirror simply floats.
For a wide but thin box, the critical acceleration will depend only on
the mass density (per unit volume) of the box.

Despite appearances, the existence of such an effect would 
be consistent with energy conservation, from both the inertial and
accelerated perspectives \cite{UW}.  Nevertheless, it is hard to see how
it could fail to lead to a disturbing instability of the vacuum and the
possibility\footnote%
{As the box passed $a=a_c$, 
 the ``push'' required to keep it accelerating
 would change into a ``pull'' (corresponding to a negative net inertial
 mass).  Thus a box held at some $a>a_c$ could deliver an infinite
 amount of work.}
of perpetual motion machines.  Thus, self-acceleration seems unphysical,
and we have referred to this issue in our title as ``the
self-accelerating box paradox.''

We address this paradox in section \ref{LIDOF} below, where we consider
a specific type of mirror that operates through the use of light
internal degrees of freedom.  For boxes with mirrored walls of this
type, we resolve the self-acceleration paradox by showing that the
mirrors necessarily heat up and in doing so acquire enough mass to
overwhelm the ``buoyant force'' on the box, no matter how great the
acceleration.  The danger of vacuum instability and perpetual motion
machines from this quarter thus seems to be removed.

However, it was the ``floating box'' effect that was used in \cite{UW}
to argue that the generalized second law could hold without the
imposition of novel entropy bounds \cite{Bek73} on matter fields.  We
therefore reconsider this issue in section \ref{GSL}.  Although this
section is motivated by our study of the self-accelerating box paradox,
it constitutes a logically independent discussion that in no way relies
on the results of the earlier sections.  We find that a derivation very
similar to that of \cite{UW} can be carried out up to a point (as long
as the other assumptions made there are retained).  However, at depths
comparable to what would have been the box's ``floating point'' the
picture changes and one is forced to consider a regime in which
``thermal radiation'' is dominated by things like box-antibox pairs.
This region is a plausible source of effects through which the second
law might be maintained, but we are unable to reach a definite
conclusion on this question.

\section{Why mirrored boxes don't float -- some models and a conjecture}

\label{LIDOF}

We believe that the key to the self-acceleration paradox lies in a study
of the mirrors themselves.  If this is so, then physical models of the
mirrors are needed.  Let us therefore consider a mirror that reflects
radiation of a certain type, corresponding to excitations of a specific
``bulk'' field $\phi$.

One can imagine two mechanisms by which such a mirror could function.
The first is that the mirror could represent some sort of fundamental
potential for $\phi$.  This might happen, for example, if the mirror
were itself a domain-wall soliton of the field $\phi$, in which case the
mass of the linearized perturbations $\delta\phi$ would in general
depend on position relative to the wall.  Under certain circumstances,
the scattering of $\delta\phi$ off of the resulting potential might be
strong, making the domain wall act as a good mirror.  For lack of a
better term, we shall refer to such mirrors as {\it solitonic mirrors}
in the discussion below.

Note, however, that the mirrors encountered in daily life are decidedly
not of this sort.  Instead, a standard silver mirror functions by a
rather different mechanism.\footnote%
{Though in the end it might be that the interaction can still be
 described via an effective, frequency dependent potential.}
When a light wave is incident on a silver mirror, the mobile electrons
in the mirror rapidly adjust their configuration so as to cancel out the
incoming electric field.  The result is that the motion of the
conduction electrons conspires to send radiation back in the direction
from whence the original signal arrived.  Because the rapid
reconfiguration of these electrons is essential, they necessarily
represent ``Light Internal Degrees Of Freedom'' of the mirror, and we
will call mirrors that function along these lines {\it LIDOF mirrors}.

Of course, one could also imagine mirrors that operate via a combination
of these two mechanisms.  One might also ask whether there could be some
third mechanism through which a mirror could function.  While we are not
able to rule out this possibility, it is hard to imagine what this third
mechanism might be.  In the other direction, one could ask whether 
one might construe the
first mechanism as a special case of the second.

In the rest of this section we shall address the familiar category of
LIDOF mirrors and show that they necessarily become heavy when placed in
contact with a thermal bath.  Due to their many internal degrees of
freedom, they will absorb a significant quantity of heat from their
environment so that their internal energy and their weight in a
gravitational field will both increase with temperature.  We will see
that this effect is strong enough to overcome any buoyancy force from
the radiation fluid, whence the mirror cannot float.  Thus,
self-acceleration of such a mirror will not occur.
(While solitonic mirrors are not addressed in this section, they will be
the subject of several comments in section \ref{disc}.)
Although our concern in this section will be with self-acceleration in
Minkowski spacetime, we note that almost all of our considerations will
apply 
{\it mutatis mutandis} 
to the case of a (small) box suspended in
the vicinity of a black hole. 

\subsection{A conjecture and some evidence}

The basis for our analysis of LIDOF mirrors will be the following
conjecture: 
\begin{quotation}
\noindent
A perfect LIDOF mirror requires at least one internal field degree of
freedom for each bulk degree of freedom that it will reflect.
Specifically, the internal degrees of freedom of the mirror must be
approximately described as a set of fields in a spacetime of one
dimension less than that of the bulk, and these internal fields must
have a minimum of one helicity state for each helicity state of the bulk
fields.
\end{quotation}
The emphasis here is on the number of internal degrees of freedom
required.\footnote%
{The counting here is of {\it effective} degrees of freedom.  Such
 degrees of freedom include, for example, those of weakly interacting
 quasiparticles describing the excitations of an underlying set of
 fields or particles which themselves could be strongly coupled.}
While we have no proof of this conjecture, the following
gedankenexperiment provides 
a certain amount of support.

Consider the scattering of an electromagnetic wave by an electron.  One
description of this process is that the wave excites the electron, after
which the electron re-radiates the wave in a different direction.  With
enough electrons and enough scattering, one can effectively create a
mirror.  Now if we naively count degrees of freedom we find two velocity
components for the electron gas, matching the two helicity states of the
reflected photons.  Moreover, if we imagine that the electrons reside in
a superconductor instead of a normal metal, we trade the velocity vector
for a scalar ``Higgs'' field describing the condensate, but since it is
a complex scalar, we again recognize two degrees of freedom per point.

Suppose now that the electron couples not just to a single
electromagnetic field but to two such fields, EM1 and EM2, and suppose
for simplicity that it has the same charge for both.  Then our mirror
for EM1 will also act as a mirror for EM2.  However, consider what
happens when a wave of EM1 and a wave of EM2 reach the electrons at the
same time.  If we arrange for these waves to be exactly opposite in
sign, their effects on the electrons will cancel and the electrons will
remain undisturbed.  As a result, the electrons will not react back on
the fields and the waves will pass through the mirror unhindered.  Said
differently, since the electrons have the same charge for both fields,
they couple only to the symmetric combination of EM1 and EM2.  The
anti-symmetric combination does not interact with the electrons at all,
whence the mirror is transparent to its excitations.
This would appear to be a general rule: coupling additional gauge fields
to a given set of charges results in the charges decoupling from certain
linear combinations of the new set of fields, with the result that no
more independent linear combinations couple to the charges than
originally.  
Therefore, one cannot violate the conjecture in this way.  However, more
subtle effects remain to be ruled out.

\subsection{Weight and buoyancy in thermal equilibrium}

We have argued above that a LIDOF mirror must have a significant number
of internal degrees of freedom, which can, at least effectively, be
described as fields living in the worldvolume of the mirror.  Since
these mirror fields are responsible for reflecting the bulk fields, they
must interact strongly with the latter, and this should be enough to
allow them to come into thermal equilibrium with 
the bulk fields.\footnote%
{For a worked out example of such thermalization in a simple case see
 \cite{caldleg}.  Notice that no self-coupling of the internal degrees
 of freedom is involved in this example.  Indeed, there is only one such
 degree of freedom --- that of a simple harmonic oscillator --- and its
 self-coupling (nonlinearity) is zero.}
Thus, when the mirror is placed in a thermal bath, one would expect the
relevant internal degrees of freedom to thermalize on a short 
timescale, with the result that
the mirror will absorb heat if initially it was cold.

A possible objection to this conclusion might be that there might exist
types of scattering analogous to the M{\"o}ssbauer effect.  However, we
will be considering only boxes whose transverse dimensions are much
greater than a wavelength $\lambda$ of the thermal radiation they are
reflecting.  In this case, it should be possible to think of the
radiation as composed of wave packets much smaller than the box wall
reflecting them, and causality forbids such a packet from bouncing off
the box as a whole.  We would thus expect that localized excitations of
the relevant internal field must result.  (In the opposite case of boxes
much smaller than $\lambda$, causality imposes no such restriction, but
that case is anyway not amenable to the analysis of this paper, because
the fluid approximation we will be using becomes invalid.)

We have argued that the internal degrees of freedom of a perfectly
reflecting LIDOF mirror must evolve rapidly to a temperature equal to
that of the bath with which they are in contact.  We will now use this
property to derive a lower bound on the energy density 
of a perfect mirror
in contact with a heat bath.  This excess energy is a function of the
ambient temperature, and we will show it to be sufficiently great that
LIDOF mirrors can never float or self-accelerate.

Given that the mirror helicity states must be in correspondence with the
bulk helicity states, it clearly suffices to treat one helicity state at
a time, because the full heat absorbed by the mirror is just a sum over
mirror helicity states, and the full ambient pressure is a sum over the
corresponding bulk helicity states.

In this section we will consider the case of massless fields, though
some comments on the massive case will be made in section \ref{disc}.
Moreover, we limit ourselves throughout to free (or almost free) fields,
both in the mirror and in the ambient spacetime.  (Of course the
interaction between bulk and mirror fields is not taken to be weak.) 
Finally, we work in the adiabatic limit, so that the bulk and mirror
fields will always be in equilibrium\footnote%
{More generally our analysis will apply whenever the equilibration time
 is sufficiently short.  If we are right about the rapid thermalization,
 this is not much of a restriction in practice.  In any case,
 consideration of the adiabatic limit is enough to rule out ``floating''
 self-acceleration, which would be a steady state condition.}.
Considering a co-dimension one mirror in a $d+1$ dimensional bulk
spacetime, we are then left with a rather simple calculation involving
the statistical mechanics of a $(d-1)+1$ dimensional massless field and
a d+1 dimensional massless field.

Let us first consider the heat absorbed by the mirror.  We will take the
mirror to be much larger than the ambient thermal wavelength, so that we
may use a fluid approximation in computing the energy density of a
thermally excited internal field.  Notice that, since we are treating
this field as a $(d-1)+1$ dimensional system, the thickness of the
mirror 
will
be small compared to the thermal length scale --- it
is only the lengths along the mirror face that must be large.

In this approximation, the energy density of a $(d-1)+1$ dimensional
massless single-helicity bosonic field is given by the expression (see,
e.g. \cite{Huang} for the 3+1 case):

\begin{eqnarray}
\label{rho}
 \rho_{d-1} 
 &=& 
 \frac{1}{(2\pi)^{d-1}} 
 \int d^{(d-1)}k 
 \frac{\omega_k}{e^{\beta \omega_k -1}} \cr
 &=& 
 \frac{V_{d-2}}{(2\pi)^{d-1} \beta^{d}} 
 \int_0^\infty dx 
 \frac{x^{d-1}}{e^x-1},
\end{eqnarray} 
where $\beta = 1/T$ and we use units in which Boltzmann's constant and
the speed of light are unity.
Here, 
$V_{d-2}=2\pi^{(d-1)/2}/((d-3)/2)!$ 
is the volume of the unit $d-2$ sphere
and
$\omega_k$ is the energy of a particle with wavenumber $k$.
For our massless field we have, of course, $\omega_k=|k|$,  and we have
used this in the second line above, having also performed
the change of variable $x=\beta k$.

Suppose that we wish the mirror to have a proper acceleration $a$, which
is related to the ambient temperature through \cite{U} $a = 2 \pi/\beta$.
Then the mirror needs to experience a co-moving force per unit area of
\begin{equation}
\label{weight}
 F/A 
 = a \rho_{d-1} 
 = \frac{V_{d-2}}{(2\pi)^{d-2} \beta^{d+1}} 
   \int_0^\infty dx \frac{x^{d-1}}{e^x-1} .  
\end{equation} 

Now, what is the actual ``buoyant'' force per unit area on the box
supplied by the ambient thermal radiation?  In discussing this question,
we will regard as ``downward'' the direction of the gravitational 
force
felt by the box in its own rest frame.  In the case of a box suspended
in the vicinity of a black hole horizon this terminology has a literal
meaning, but it is also convenient for the case of 
concern here ---
that of a box accelerating in a flat background spacetime.  Now, the
buoyant force is of course just the difference in pressure between the
top and bottom of the box, which is bounded above by the ambient
pressure on the mirror forming the bottom of the box.\footnote%
{Since this bound ignores the pressure on the top of the box, our
 analysis applies not only to a box lowered through a region in thermal
 equilibrium, but also to upward directed thermal radiation of the sort
 that is present in the Unruh ``vacuum'' near a black hole.}
Strictly speaking, this pressure should be calculated from the
renormalized stress energy tensor of a quantum field, with appropriate
boundary conditions imposed at the walls of the box.
However, since such results are not presently
available we shall instead follow the approach of \cite{UW} and assume
that the ambient pressure is in fact equal to that of a thermal fluid in
an enclosure of large volume.

Bekenstein has argued \cite{buoy} that the pressure is somewhat less
due to inefficiencies in the scattering of finite wavelength radiation
by the box.  There may also be 
other ``finite size effects'' which are
missed in the way we approximate the pressure.\footnote%
{Readers concerned about finite size effects \cite{LL} due to the fact
 that, even in flat spacetime, the scale over which the temperature of
 the Unruh radiation changes is comparable to a thermal wavelength
 may consult \cite{WA}, which provides some evidence that such effects
 may be ignored.
 In the corresponding black hole case, there might be further ``finite
 size'' effects
 that we have not accounted for arising from the spacetime curvature.}  
However, since our goal is simply to remove an effect deduced in
\cite{UW}, it is consistent to use their fluid description even if
this somehow leads to an overestimate of the ambient pressure.

In this approximation, the pressure (for free fields, as always) is
given by an integral much like (\ref{rho}) and we have

\begin{eqnarray}
\label{P}
  P_{d} &=& \frac{1}{d(2\pi)^{d}} \int d^{d}k 
  \frac{\omega_k}{e^{\beta \omega_k -1}} \cr
  &=& \frac{V_{d-1}}{d(2\pi)^{d} \beta^{d+1}} 
  \int_0^\infty dx \frac{x^d}{e^x-1} .
\end{eqnarray} 

Our task is now to compare the buoyant force to the weight of the box.  
{}From (\ref{weight}) and (\ref{P}) their ratio is
\begin{equation}
\label{ratiobw}
  {1 \over (2\pi)^2} {V_{d-1}\over V_{d-2}} {f(d)\over f(d-1) d} \ ,
\end{equation}
where we have introduced the definition
\begin{equation}
\label{1}
  f(d) = \int_0^\infty dx \frac{x^d}{e^x-1} \ .
\end{equation}
We claim that this ratio is less than $1/4\pi$ for all $d>1$,
reducing in particular to $\pi^2/180\zeta(3)=0.0456$ for the most
important case of $d=3$.  In other words the buoyancy is more than an
order of magnitude smaller than the weight in all dimensions and more
than 20 times smaller in $3+1$-dimensions.

In fact this is not hard to see, using the known formulas
$$
   V_d =  2 \pi^{d+1\over 2} / ({d-1\over 2})!
$$
and
$$
   f(d) = d! \, \zeta(d+1)  \,
$$
where $\zeta$ is the Riemann zeta-function:
$$
       \zeta(s) = \sum_{n=1}^{\infty} {1\over n^s}\ .
$$
(The formula for $f(d)$ can be derived by expanding the denominator in
(\ref{1})
and noting that $\int_0^\infty dx x^n e^{-x} = n!$.  The formula
for $V_d$ follows directly from the formula $\pi^{d/2}/(d/2)!$ for the
volume of the unit $d$-ball (or ``disk''), which in turn can be looked
up or proven by induction.)

Substituting this formula for $f(d)$ converts (\ref{ratiobw}) into
$$
  {1 \over (2\pi)^2} 
  {V_{d-1}\over V_{d-2}}
  {\zeta(d+1)\over\zeta(d)}
$$
and the claimed bound then follows from the obvious inequality
$\zeta(d)>\zeta(d+1)$ 
and the easily checked bound 
$V_d/V_{d-1}\le\pi$ for $d\ge{1}$. 

Thus, in all cases we find that the ambient pressure is insufficient to
maintain the acceleration of the box.

\section{Black Holes and the Generalized Second Law}
\label{GSL}

Given that a LIDOF mirrored box fails to float in any bath of thermal
radiation, it seems worthwhile to reconsider the Bekenstein-Unruh-Wald
discussion of whether the laws of 
thermodynamics, as applied to black
holes, impose novel entropy bounds on the matter in the universe (say of
the form in \cite{Bek73}).  In \cite{UW}, Unruh and Wald used the
floating box effect in analyzing the implications of the generalized
second law for a process in which a perfectly reflecting box is lowered
adiabatically from infinity to some location near the horizon and then
dropped into the black hole.  They found that the second law was exactly
saturated if the box was dropped just at its floating point.  With that
choice of release point, the mass contributed to the black hole by the
box augmented the black hole's entropy by an amount that exactly offset
the disappearance of the entropy contained in the box.

However, since, according to our analysis, the box absorbs heat from the
thermal bath and thus increases its weight, one can extract more energy
during its descent than in the Unruh-Wald analysis.  As a result, by
lowering the box adiabatically and releasing it at the point at which
the corresponding Unruh-Wald box would float, one might expect to be
able to decrease the net entropy of the universe, in contradiction to
the generalized second law of thermodynamics.\footnote
{As usual in these discussions, one means by ``entropy of the universe''
 the sum of all entropy external to the black hole(s) with the entropy
 belonging to the horizon area of the black holes themselves.  In
 particular, one does not count explicitly the entropy of any object
 that has fallen through a horizon.}

In opposition to any such scenario, reference \cite{GW} points out that
if the black hole system can be described as a standard state of thermal
equilibrium then no violation of the second law can occur, regardless of
the detailed behavior of any perfect mirrors.  However, this begs the
question of whether the posited description is appropriate and, if it
is, of whether the usual expressions for black hole entropy are
consistent with this description in the absence of novel entropy bounds.
Nor does it tell us what prevents the box from being lowered to its
erstwhile floating point as envisioned above with the extraction of more
energy than the second law can tolerate.  

In the following, we explore to what extent the analysis of \cite{UW}
can 
be preserved given the failure of boxes to float.  Let us
first consider the entropy balance while the box is being
quasi-statically lowered.  In this phase of the process, the only change
in entropy is that associated with the heat flow from the thermal
atmosphere of the black hole into our mirror.  Since heat will flow only
in a direction that generates entropy, this should create no
difficulties with the second law.

It remains to analyze the second stage in which the box is dropped into
the black hole.  Just before this occurs, the box has some total energy
$\epsilon$ and the black hole some total mass $M$.  (Here, the energies
$\epsilon$ and $M$ are measured relative to the timelike Killing field
with unit normalization at infinity.)  We remark that $M$ at this stage
may be smaller than when the box was far away, due to the energy that
flowed from it, through the thermal atmosphere, and into the box during
the lowering process.  Similarly, $\epsilon$ includes the heat that
flowed into the box while it was being lowered along with any other
energy that might have entered the box.

Assuming that the box carries no charge or angular momentum, the first
law of black hole mechanics tells us that dropping it into the black
hole increases the latter's entropy by an amount
\begin{equation}
\label{ratio}
    \delta S_{bh} = \frac{\epsilon}{T_H} = \frac{E}{T} 
\end{equation}
where $T_H$ is the black hole's temperature relative to infinity and $E$
and $T$ are the {\it locally} measured energy of the box and locally
measured temperature of the thermal radiation.  Note that $E$ and $T$
are related to $\epsilon$ and $T_H$ by the same redshift factor, which
therefore cancels in the ratio (\ref{ratio}).

The locally measured energy $E$ consists of two pieces.  The first,
which we shall call $E_{box}$, is the energy that the particular
internal state of the box would possess in isolation.  But, since it
requires work $PV$ to bring a box of volume $V$ into the high pressure
($P$) environment of the thermal atmosphere, the locally measured energy
has a second contribution, and is given in full only by the sum
$E=E_{box}+PV$.

Taking into account the entropy $S_{box}$ that disappears into the black
hole, the total change in entropy is thus
\begin{equation}
  \delta S 
  = \delta S_{bh}         - S_{box} 
  =  \frac{E}{T}          - S_{box}   
  =  \frac{E_{box}+PV}{T} - S_{box}.
\end{equation}

We now make use of the Gibbs-Duhem relation, $PV=-E_{rad}+TS_{rad}$ as
in \cite{UW}, 
which relates the pressure and volume of thermal
radiation to its energy $E_{rad}$ and entropy $S_{rad}$.  Under the
assumption that the radiation constitutes a {\it homogeneous and
isotropic fluid} whose entropy and energy are additive, this relation
can be derived by integrating the first law $PdV=-dE_{rad}+TdS_{rad}$ at
constant pressure and temperature.\footnote
{Near the horizon conditions are far from homogeneous and isotropic.
 Rather the thermal wavelength of massless radiation is comparable to
 the scale on which the locally measured temperature varies.  For this
 reason, one may be skeptical of using the Gibbs-Duhem relation in the
 present context.  On the other hand, the results of \cite{WA} encourage
 the belief that the pressure at a certain depth does in fact correspond
 to that of an unconfined thermal fluid at the appropriately redshifted
 temperature.  As energy conservation relates the density to the change
 of pressure with depth, the energy density would also be that of an
 unconfined thermal fluid.  This would be enough for the Gibbs-Duhem
 relation to hold.}
(One is also assuming that the ``chemical potential'' is zero in the
sense that no conservation law freezes the abundance of any constituent
of the fluid.)  Substituting this result in the above equation, we find
\begin{equation}
\label{compF}
   \delta S 
  = \left(\frac{E_{box}}{T} - S_{box} \right) 
  - \left(\frac{E_{rad}}{T} - S_{rad} \right) 
  = \frac{F_{box} - F_{rad}}{T} ,
\end{equation}
where $F_{box}$ denotes the Helmholtz free energy of the box and $F_{rad}$
that of an equal volume of thermal radiation at temperature $T$.

We now recall that thermal radiation minimizes the free energy at fixed
temperature\footnote%
{Of course, this remark would be irrelevant if the ``thermal
 atmosphere'' of a black hole were not truly described by a thermal
 ensemble in this sense, i.e. by a density operator of the Gibbs form
 $e^{-\beta H}$ which, by the usual arguments of statistical mechanics,
 is the one that minimizes the free energy at fixed temperature.
 Evidence that this density operator correctly describes the fields
 surrounding a black hole comes from the results of
 e.g. \cite{SH,FH,rdsse} and especially \cite{UW2,Ray,BFZ,TJ1,TJ2,TJ3}
 which indicate that the Hawking radiation is described by this density
 matrix (for the out-going modes).  There is also strong evidence from
 \cite{BisWich}, \cite{smbhthd} and \cite{FH}
 that any state of the quantum fields
 in a fixed black hole geometry (and contained in surrounding walls)
 will evolve toward the Gibbs state possessing the Bekenstein-Hawking
 temperature of the black hole.}.
If we assume that the presence of the box does not disturb the
free-energy density of the surrounding radiation, and that one can
neglect the variation of the latter between the top and bottom of the
box, we can 
conclude 
that
$F_{rad}<F_{box}$ and therefore that $\delta S$ must be positive.  

To obtain a physical picture of how this may happen, it is helpful to
recall that the free energy of an object controls its probability 
to occur
in a thermal ensemble.  In other words, the effective quantum
field that describes boxes such as ours will be thermally excited at
temperature $T$ just as will any other quantum field near the black
hole.  When the region of space at temperature $T$ is large compared to
the size of the box, then $F_{box}$ should control the probability for
boxes to appear from thermal fluctuations.

What we may conclude from this is not that $F_{box}$ is somehow bounded
below by some bound set by $F_{rad}$.  
Instead, the implication is that, 
were $F_{box}$ to
become sufficiently negative, the thermal radiation would consist
primarily of box-antibox pairs (as well as related flotsam like
fragments of box 
walls) rather than ordinary massless quanta.
Unfortunately, one's understanding of such a ``fluid'' is minimal, and
we have no real grounds on which to judge how well the assumptions we
have had to make in the above analysis survive into this new regime.  In
particular it seems an open question whether anything like the
Gibbs-Duhem relation would continue to hold.  
Thus, at this stage we are unable to 
demonstrate that 
the second law is maintained without a better understanding of
the ``fluid of boxes and box fragments.''  We will return to 
these questions in the discussion section.

Notice 
that the discussion in this section was completely
independent of our LIDOF mirror model and should apply to solitonic box
walls as well.  However, this in itself sheds no light on the extent to
which our resolution of the self-acceleration paradox carries over to
the case of solitonic mirrors.

\section{Discussion}
\label{disc}

In this paper, we have addressed two separate but related issues 
stemming from consideration of 
classic thought experiments
concerning the generalized second law of thermodynamics.  The first was
the question of whether perfectly reflecting boxes can self-accelerate
in flat spacetime (which in essence is equivalent to the question of
whether such boxes can float \cite{UW} in the thermal atmosphere of a
black hole.)  Because the existence of such ``runaway solutions''
would conflict 
with cherished beliefs like the stability of the vacuum and
the impossibility of a ``free lunch'', one might expect that they cannot
actually occur.

In addressing this paradox, we limited ourselves to mirrors that operate
through Light Internal Degrees of Freedom (LIDOF mirrors) and suggested
that such mirrors will require a minimum of one internal field degree of
freedom (helicity state) for each bulk field degree of freedom (helicity
state) to be reflected.\footnote%
{In addition the box wall should have at least one acoustic mode, and
 this mode should also come into thermal equilibrium with any ambient
 heat bath, for the same reasons of causality as for the other modes.}
Although we did not prove this assertion, we gave a suggestive example
along with supporting arguments.

Accepting this conjecture, we demonstrated that the heat-energy absorbed
by such mirrors is more than sufficient to prevent any self-acceleration
resulting from the coupling to massless fields in flat spacetime.  It is
interesting to note that our calculation leaves significant margin for
error in the conjecture.  For example, in 3+1 dimensions our conclusions
would survive even if the internal degrees of freedom were down by a
factor of 20.  (It is also clear that one could weaken the claim by a
factor of $\sqrt{d}$ and still achieve the desired result.  In fact,
numerical calculations show that one could weaken the inequality by a
factor of the spatial dimension $d$ and that the result would still hold
up to $d$ on the order of a few hundred.)

In addition to the question of the validity of our conjectured bound on
the number of internal degrees of freedom of a LIDOF mirror, several
other questions are raised by our discussion.  One of them concerns the
other category of mirrors (``solitonic mirrors'') and whether one can
also derive general constraints forbidding their self-acceleration.
While we are not certain how this would work, it seems plausible that a
soliton which provides a strong reflecting potential for some field must
itself incorporate a strong potential which in turn would force it to
have a large tension (and therefore a large mass).  Another idea is that
the effective description of the soliton in terms of an effective
potential might be concealing light internal degrees of freedom which
come into play in the course of reflecting external fields.  A clear
first step in investigating either of these ideas would be to study
particular model systems.

A second question concerns the case of massive fields, which are
significantly harder to excite when placed in contact with a heat bath.
How would the buoyancy compare with the weight if massive fields were
the only relevant ones (eg because there were no massless fields in
nature or because our mirror was transparent to them)?  Of course, if
the mass is small compared to the thermal energy ($m \beta \ll 1$) then
such fields will behave as if they were massless.  However, for fields
with masses higher than the thermal energy scale the situation is rather
different.  Considering a bulk field of mass $m$ and a mirror field of
the same mass\footnote%
{One expects that a mirror field of greater mass would be unable to
 adjust sufficiently quickly to reflect the bulk field.  On the other
 hand, making the mirror field lighter than the bulk field would only
 strengthen our case.},
one finds that for $m\beta \gg 1$ the ambient pressure is
\begin{equation}
\label{massive}
   P \sim e^{-\beta m} \frac{m^{d/2}}{\beta^{(d+2)/2}},
\end{equation}
whereas the heat absorbed by the mirror adds a weight per area of
\begin{equation}
  \frac{W}{A} 
  \sim e^{-\beta m} \left( \frac{m}{\beta} \right)^{(d+1)/2}
  \sim P {\sqrt{m \beta}}   \gg P .
\end{equation}
We see that, in this case also, the additional weight is more than
sufficient to prevent self-acceleration.

Some readers may feel that our analysis does not just place bounds on
perfect mirrors, but shows instead that perfect mirrors do not exist at
all.  They may point out that if the mirror thermalizes with the
exterior, it must also radiate thermally into the interior (and vice
versa).  Thus, the mirror couples the inside with the outside.  This is
of course true to some extent, but one did not need to count the
internal degrees of freedom to make this argument.  The relevant
question is whether the timescale for this energy flow can be made
arbitrarily long.  Certainly, our arguments require that a minimum
number of degrees of freedom interact with the bulk fields on a short
timescale.  That they need not interact equally quickly with the
interior of the box can be made clear by considering a layered mirror
composed of several copies of the simple mirrors studied above.  The
timescale for transmitting energy between interior and exterior will
grow without bound as more and more mirrors are added to the
laminate.\footnote
{Note, however, that the next few layers of our multi-layer mirror will
 also interact quickly with the bulk fields and thus absorb even more
 heat from the thermal bath.}

In Section \ref{GSL}, we considered the implications of our earlier
conclusions for the generalized second law discussions conducted by
Bekenstein \cite{Bek73} and Unruh-Wald \cite{UW}, which historically
were the source from which the self-accelerating box paradox arose.
Since the main point at issue in those discussions was the relative
importance of the buoyant force on a mirrored box, one might think that
our results would modify the terms of the debate, and in that way
perhaps resolve it.

In the event, however, it seems that the former has occurred without the
latter.  In particular, we constructed a variant of the derivation of
\cite{UW} which arrives at similar conclusions without assuming that the
box will float.  Our derivation is independent of section \ref{LIDOF}
and therefore applies equally well to solitonic mirrors as to LIDOF
mirrors.  However, we now recognize that the interesting regime is one
in which the boxes themselves become an important constituent of the
thermal atmosphere.

Let us reflect in more detail on what we have learned from these latter
calculations.  Some boxes, such as those satisfying the entropy bound
$S/E < 2\pi R$ of \cite{Bek73}, were long ago established to be `safe'
in the sense that lowering such a box into a black hole respects the
second law even when the thermal atmosphere of the black hole is
ignored.  In our current language such boxes always have $F_{box}>0$
since their large size keeps the box far enough from the horizon that we
have $T \le (2 \pi R)^{-1}$. Our calculation once again finds that such boxes
are safe and, furthermore, that the existence of such boxes is
consistent with any additional assumptions about the black hole's
thermal atmosphere so long as $F_{rad}<0$ (which in turn is equivalent
to positivity of the pressure if the Gibbs-Duhem relation holds).  The
assumption that the atmosphere (and in particular the contributions of
boxes and anti-boxes) can be neglected amounts to taking $F_{rad}=0$.

However, if we imagine a box that violates the above entropy bound then,
as it descends toward the horizon and the ambient temperature increases,
one may indeed enter a region where $F_{box}<0$.  In this case,
neglect of the thermal atmosphere would lead to an apparent violation of
the second law.  Similarly, suppose one makes some other assumption as
to the nature of $F_{rad}$.  For example, one might assume that
$F_{rad}$ is dominated by the contributions from a certain class
${\cal{C}}$ of states, leading to a relation of the form $F_{\cal C} =
F_{rad}$.  If the Gibbs-Duhem relation holds for this class of states,
(\ref{compF}) requires the restriction $F_{box}>F_{\cal C}$ for the
generalized second law to hold.  However, since the free energy controls
the relevance of an object or class of states to a thermal ensemble, the
existence of a region with $F_{\cal C}>F_{box}$ implies simply that
thermal fluctuations in this region include box anti-box pairs in
significant numbers and that the class of states ${\cal C}$ does not in
fact dominate their contribution to the free energy.

When a box enters such a region several new occurrences would seem to be
possible which might (or might not) intervene to protect the second
law.  First, there might be a significant probability that our box would
be destroyed by collision with an anti-box.  If this happened, we could
extract no further energy by trying to lower it adiabatically.  Second,
with boxes being produced thermally we might observe one being emitted
from the black hole -- and when combined with the destruction of our own
box by an anti-box we might well interpret this as our box ``bouncing
off'' the thermal atmosphere.

Of course, such effects would not be needed if the analysis of Section
\ref{GSL} could be taken over unchanged to the new regime.  The key
point here is that if boxes now dominate the thermal radiation, we have
$F_{box}=F_{rad}$ as an identity, whence $\delta S$ would vanish
identically even if it were somehow meaningful to lower the box to this
depth and drop it into the black hole.  In this way, one would again
maintain the generalized second law without appealing to novel entropy
bounds.  However, it is an open question\footnote%
{Most of these uncertainties relate to ``finite size effects'' in the
 broad sense of the term.  See for example the comments in \cite{buoy}
 about the effects of finite wavelength on the scattering of thermal
 radiation.  And no one who harbored such doubts before is likely to be
 any happier when they are told that now one is talking about a fluid of
 box-antibox pairs rather than photons.}
whether the fluid model we used in Section \ref{GSL} could be adequate
for computing the ``buoyant force'' on the box in a domain where boxes
(and box fragments) become a dominant component of the thermal
radiation.\footnote%
{More generally, the critical question is whether the thermal atmosphere
 can provide sufficient buoyancy to insure that the work gained by
 lowering the box (including its mirrored walls) is not so great as to
 exceed the box's initial entropy appropriately weighted by the black
 hole temperature.}
We used such a model in at least three ways: to justify the
``work'' term $PV$ in (3.2); to justify using the Gibbs-Duhem relation,
which depends on the local homogeneity and isotropy of a fluid; and to
justify the assumption that free energy changes are additive when one
replaces the box by an equal volume of thermal radiation.\footnote
{Notice also that, even for more ordinary thermal radiation, the fact
 that (in the regimes of interest) the box is never more than one or two
 thermal wavelengths from the horizon calls into question the adequacy
 of ignoring the inhomogeneities and anisotropies induced by boundary
 conditions at the box walls.}

The status of entropy bounds (in the sense of universal bounds that
would hold in an arbitrary theory) thus remains elusive.  There seems to
exist neither a conclusive argument that they must hold nor a conclusive
argument to the contrary.\footnote%
{It used to be thought that a world with exponentially many species $N$
 of particles/fields could serve as a counterexample (eg \cite{zhang}),
 but this also can be questioned, at least in the case of an area bound
 like $S<A/G$, as it neglects the $N$-dependent renormalization of
 Newton's constant $1/G$.  
 As envisioned in connection with entanglement entropy \cite{tangle} 
 by \cite{lenug} and \cite{ted}, this correction would, 
 if the arguments of \cite{lenug} are correct, 
 more than compensate the contribution of a large $N$ to $S$.
 See, however,
 \cite{sol} and \cite{kabat} for comments on the ideas of \cite{lenug}.}
The main contribution of the above analysis in this connection is
probably to demonstrate that the question cannot be decided without
consideration of a putative new regime in which thermal ``radiation''
would no longer be dominated by massless fields.\footnote%
{Notice that the analysis in Section \ref{LIDOF} of whether a mirrored
 box can self-accelerate/float would also have to be reopened in such a
 regime.}
In opposition to earlier suggestions that (in the context of adiabatic
box lowering) the second law necessarily fails in the absence of entropy
bounds, we have suggested plausible mechanisms for maintaining the
second law in this new regime.  On the other hand we certainly have not
analyzed these mechanisms fully and it would not seem to be an easy task
to do so.

The attentive reader 
may point out
that the discussion in \cite{UW} showed
the second law to be only marginally satisfied when the box
was filled with thermal radiation.
Since more
energy is removed from the system in our scenario than in that of
\cite{UW}, the energy of the box should be less at any height and, if
the box is dropped from the putative floating point, less energy
would be transferred to the black hole.  Thus, 
one might wonder how we can avoid second law violations for this case.
Clearly, to raise this objection one must tentatively assume that
the thermal radiation mentioned above does not include significant
numbers of box/anti-box pairs.

However, without the inflow of heat into the walls, our box was chosen to be
indistinguishable from thermal radiation at the putative
floating point.  Thus, without this inflow, it is exactly at this
point that $F_{box}=F_{rad}$ would first be achieved.  The
issue just raised may be resolved if the net effect of the heat inflow
is to lower the free energy of the box, so that this transition occurs
higher in the black hole's thermal atmosphere. 

To see that this is the case, we divide the energy $E_{box}$ into $E_0$, the
locally measured energy of the box before it was lowered toward the
black hole and $Q$, the total heat that flowed into the box while being
lowered.  We further write
$Q = \int q(z) \ dz$ where $q(z)$ represents the locally measured heat
that flowed into the box between depths $z$ and $z+ dz$.  Similarly, we
divide the entropy of the box into the initial entropy $S_0$ present
before the box was lowered and the entropy gain $S_Q$ resulting from the
heat absorbed by the box.  The free energy of the box  is thus
\begin{equation}
\label{divFbox}
           F_{box} =  E_0 - T S_0 + Q - T S_Q.
\end{equation}

We wish to show $TS_Q> Q$ so that $F_{box} < E_0 - T S_0$.
But since the heat flow occurred without work being done (in
particular, without changing the volume of the box), a local application
of the first law of thermodynamics yields
\begin{equation}
         TS_Q = T\int dz \frac{q(z)}{T(z)} > Q \ .
\end{equation}
In the last step we have merely used the fact that the temperature is
greatest at the lowest point yet reached by the box (i.e., the current
depth) so that $T > T(z)$.  Thus we see that for what would have been the
marginal case, box/anti-box pairs already
dominate the thermal radiation at the would-be floating point.

A final question is whether the results of Section \ref{LIDOF} above
might not be germane in analyzing some of the ``time machines'' that
people have tried to imagine.  While we know of no direct application,
many of the ``materials science'' issues that arise, for example, in
attempts to hold open ``wormholes'' are similar to those involved in
analyzing boxes with impermeable, rigid walls, like those
we have been considering in this paper.  This suggests that a very
general property of such walls, like that conjectured in Section
\ref{LIDOF}, could turn out to be important in the ``time machine''
context as well.

\begin{acknowledgments}

Both authors would like to thank Jacob Bekenstein and Bob Wald for
extensive helpful comments on an early draft of this paper and Ted
Jacobson for input concerning renormalization issues.  D.M. would also
like to thank Bob Wald and Eanna Flanagan for numerous conversations
about black hole thermodynamics and floating boxes over the past few
years.  He would also like to thank Abhay Ashtekar and Eric Poisson for
clarifying discussions.

D.M. was supported in part by NSF grants PHY97-22362 and PHY00-98747,
the Alfred P. Sloan foundation, and by funds from Syracuse University.
He would also like to thank the Aspen Center for Physics for their
hospitality during part of this work.
R.D.S. was partly supported by NSF grant PHY-0098488, by a grant from the
Office of Research and Computing of Syracuse University, and by an EPSRC
Senior Fellowship at Queen Mary College, University of London.  
He would also like to express his warm gratitude to Goodenough College,
London for providing a splendid living and working environment which greatly
facilitated the writing of this paper.

\end{acknowledgments}

\end{document}